# How much can we save? Upper bound cost and emissions benefits from commercial and industrial load flexibility

Akshay K. Rao[a], Fletcher T. Chapin[a], Erin Musabandesu[a], Adhithyan Sakthivelu[a], Carson Tucker[a], Daly Wettermark[a], Meagan S. Mauter[a,b,c,d,e †]

[a] Department of Civil and Environmental Engineering, Stanford University, 473 Via Ortega, Stanford, California 94305, United States
[b] Environmental Social Sciences, Stanford University, 473 Via Ortega, Stanford, California 94305, United States
[c] Senior Fellow, Woods Institute for the Environment, Stanford University, 473 Via Ortega, Stanford, California 94305, United States
[d] Senior Fellow, Precourt Institute for Energy, Stanford University, 473 Via Ortega, Stanford, California 94305, United States
[e] Photon Science, SLAC National Accelerator Laboratory, 2575 Sand Hill Road, Menlo Park, California 94025, United States

*†Corresponding author: mauter@stanford.edu*

## Abstract

Load shifting by commercial and industrial power consumers reduces costs and Scope 2 emissions for the consumer and the grid. Incentivizing this behavior requires tools for valuing flexibility amidst the heterogeneity in load characteristics across diverse sectors and the spatiotemporal variation in electricity prices and emissions factors. This work presents a top-down approach to screen and broadly understand the benefits of flexibility based on system uptime, power capacity (PC), energy capacity (EC), and round-trip efficiency (RTE). Depending on the region and season, cost savings from flexibility range from 0 to over 100% and emissions savings are generally bounded between 5-40%. We also find the magnitude and cost of emissions abatement from flexibility is highly variable and, in some cases, up to four orders of magnitude less than regional renewable energy credits or common investing or policy benchmarks like the social cost of carbon. While the value of flexibility is highly dynamic, estimating savings as a function of load characteristics and incentives can inform heuristic design of new systems, siting strategies, comparison of flexibility to other decarbonization options, and new avenues for incentivizing flexibility.

## Main Text

Rapid growth in U.S. electricity demand is increasing the hourly, seasonal, and yearly volatility of electricity prices and carbon emissions.[1,2] Some large load consumers (e.g., data centers, chemical manufacturing, food and beverage processing, electric vehicles, etc.) will mitigate the financial risk of this volatility by increasing on-demand generation, adding on-site storage, and delaying or foregoing electrification upgrades. Other consumers will mitigate this risk, and potentially even profit, by shifting power consumption from peak to off-peak periods by increasing the flexibility of their production schedules.[3–6]

Understanding how electricity prices and emissions intensities interact with a firm's ability to schedule power consumption is key to reliably and affordably decarbonizing both industrial manufacturing and the power grid. Stronger characterization of commercial and industrial flexibility incentives will benefit grid operators seeking to forecast peak load growth and improve grid reliability while reducing peaking generator deployment and renewable energy curtailment.[7–11] Utility commissions will benefit when seeking to use tariff design as a lever for reducing grid congestion, modernization costs, and carbon intensity.[12–16] Policy makers will benefit from understanding where investment incentives are necessary to spur electrification.[17] Finally, spatiotemporal analysis of the variance in these incentives will benefit diverse load consumers making capital upgrade decisions and innovators assessing growth markets for products that facilitate flexible load operation.[18]

Past work has primarily adopted the perspective of an individual load consumer, optimizing the timing of energy consumption for case study consumers with a defined load shape under a set of price-based or carbon-based incentive structures. Collectively, these bottom-up analyses have sampled a variety of load-shapes (i.e., system dynamics), incentive structures (i.e., price signals, emissions costs), and design upgrades (e.g., smart thermostats, variable frequency pumps, etc.).[19–24] A few case studies also impute the value of investing in flexibility upgrades,[25–29] calculate the carbon abatement potential of energy flexibility,[30,31] and quantify trade-offs between cost and emissions using marginal abatement cost curves.[32–36]

The diversity of these case studies limits their intercomparison and hinders broader insights into the interactions between consumer level attributes and grid or utility incentives that motivate flexibility. In the simplest case, consumers with large power capacities (i.e., percentage of power that can be increased or decreased above a baseline) and low uptime requirements (i.e. percentage of time the power demand can be zero) can flex their load to exploit available arbitrage opportunities. It is less clear, however, what the upper bound on the value of this flexibility might be for different grid regions. It is also unclear how loads across the spectrum of uptime and power capacity optimally respond to diverse price- or emissions-based incentive structures. For example, a system with a 100% uptime requirement that can shift a portion of its load in response to time-varying incentives (Figure 1 – blue circle) may benefit differently than a system with low uptime requirements that cannot continuously vary its power consumption (i.e., capacity) (Figure 1 – green triangle).

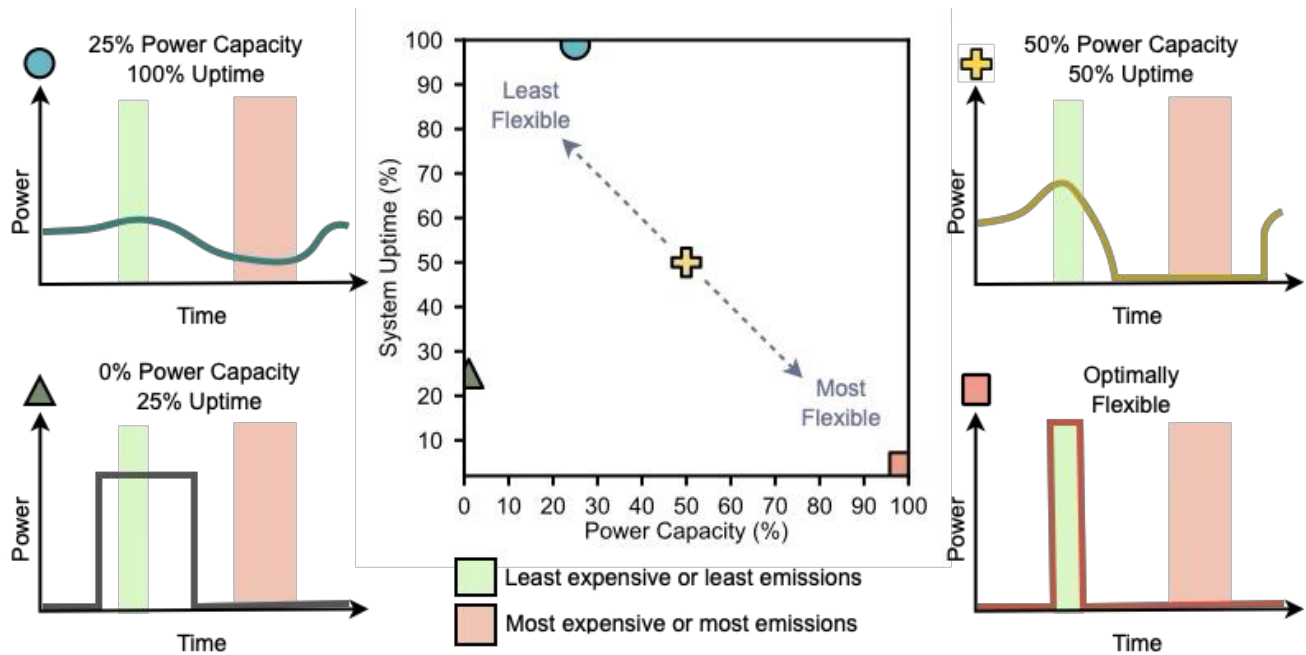

*Figure 1: Overview of system dispatchability and flexibility for example load types. Uptime refers to the fraction of time a system must be online. Power capacity refers to maximum deviation in power consumption relative to baseline or average power consumption. The most constrained system must always be online and is unable to vary its operation. The least constrained system can fully schedule when it is online, allowing for wide variation in power consumption. Load plots refer to example trajectories in which each example system might operate for a price signal that has a clear peak and off-peak incentive in cost or emissions intensity. The optimally flexible case considers 100% power capacity and the cost-optimal uptime.*

In short, there is a need for new approaches to identify which load characteristics are valuable and which cost incentives are effective for diverse power consumers. In this work, we describe a top-down approach that is generalizable, measurably optimal, and multi-objective. We value load flexibility as a function of power consumer attributes (i.e., uptime requirements), load flexibility attributes (i.e., power capacity, round-trip efficiency, and energy capacity), and regional grid attributes (i.e., marginal and average emissions factors (MEFs and AEFs) and electricity pricing via day-ahead markets (DAMs) and utility tariffs). We then formulate a mixed integer linear program (MILP) that minimizes the electricity cost or emissions of a consumer, over the power consumption schedule. The power consumption schedule has a discrete component (on or off) and a continuous component. The discrete component is constrained based on the system uptime and the continuous component is constrained based on the power capacity (PC). The total amount of energy consumed over the horizon is fixed and related to the baseload by the round-trip efficiency (RTE). The energy capacity (EC) of the system is calculated *ex post*. The analysis remains agnostic to load shape and flexibility by sweeping across uptime, power capacity, and RTE for each grid attribute and incentive. The maximum value for each consumer, flexibility, and grid attribute combination represents the upper bound of cost and emissions savings potential for different U.S. grid regions. Recent work provides data inputs for regional grid attributes.[37] Finally, we examine the cost and emissions tradeoffs from flexible operation across U.S. grid regions.

This approach allows the user to determine the value of flexibility for all feasible combinations of consumer load shape, flexibility attributes, and flexibility incentives. Doing so addresses the fundamental generalizability issue with case studies and the need for design guidelines applicable to different stakeholder groups. It also allows the user to calculate the upper bound value of flexibility in specific electricity markets, define a technology-agnostic feasibility space for flexibility-enhancing technologies with diverse PC, RTE, and EC values, and prioritize advantageous attributes for new technologies. Finally, the computational efficiency of the model enables the analysis of the tradeoffs in carbon abatement potential and cost savings benefits under different cost and carbon incentive structures.

This paper applies the approach to mechanistically understand and compare the efficacy of flexible loads. The parametric analysis informs general principles for the design for components and systems that enable flexibility in terms of these characteristics for load flexibility (e.g., should pump manufacturers focus on maximizing peak efficiency or broaden the range of operation). The broad sensitivities across grid regions can inform the siting of flexible loads to effectively exploit energy arbitrage opportunities. Multi-objective analysis uncovers where market incentives for flexibility are strong and how policy can unlock, or further incentivize, emissions reductions. Finally, we compare the order of magnitude abatement cost from energy flexibility to that of other decarbonization interventions (e.g., energy efficiency).

**Results**

We begin by identifying characteristics (i.e., uptime, PC, RTE, and EC) that minimize power consumer electricity costs and emissions under four different signals (i.e., MEFs, AEFs, DAMs, and tariffs). We plot an example using historic conditions in the California Independent System Operator (CAISO) region in July 2023 as an example (Figure 2). Next, we describe how the upper bound on cost or emissions savings varies when optimizing for each signal throughout the year and across ISO regions (Figure 3). We then identify Pareto optimal combinations of cost and emissions benefits as a function of load type (Figure 4), for the sample conditions in July 2023. From the Pareto front, we derive the cost of emissions abatement and visualize this for regions around the US with seasonal variation for a few example system types (Figure 5).

*Upper bound cost and emissions savings as a function of load characteristics*

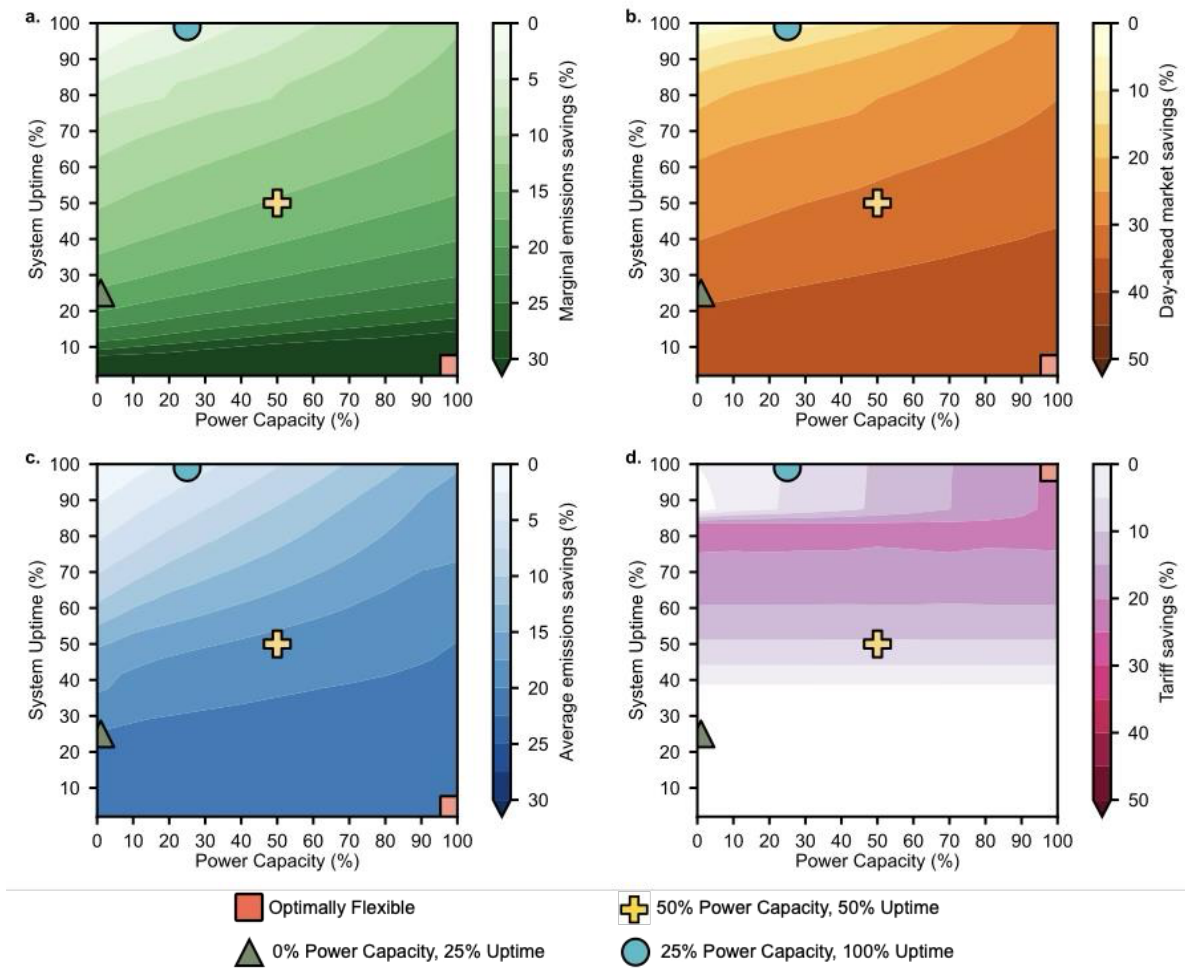

*Figure 2: Upper bound emissions and cost savings as a function of a CAISO load consumer's uptime and power capacity in July. EC and RTE are fixed at 100%. A) System operation minimizing marginal emissions. B) System operation minimizing day-ahead market electricity prices. C) System operation minimizing average emissions savings. D) System operation minimizing electricity costs under the SCE TOU-8 Option D industrial tariff. Symbols correspond to the four cases highlighted in Figure 1. Data for other months and ISO regions can be found via S1, and corresponding data is summarized in Figure 3.*

We evaluate the upper bound on cost and emissions reductions as a function of consumer load flexibility attributes (uptime, PC, EC, and RTE). In general, uptime and PC are the most measurable quantities and most strongly impact the trends in value. Since EC is difficult to measure and dependent on uptime, PC, and RTE, we fix this parameter at 100% and visualize the minimum EC for a given uptime and PC in S2. To estimate the upper bounds of value, we fix RTE at 100% but include additional sensitivity analysis on RTE in S3.

Time-invariant loads (i.e., 100% uptime, 0% PC) yield no savings relative to their baseline operation (Figure 2). For facilities with flexibility in uptime or power capacity, we find that MEF, DAM, AEF, tariff outcomes are typically more sensitive to uptime flexibility than power capacity. For the example systems in Figure 2, the tariff savings case is the only one where a system with near 100% uptime requirements (blue circle) exhibits substantial savings potential. To optimize for linear incentives (MEF, AEF, DAM, and tariffs without demand charges), it is usually preferable to have zero PC and low uptime (green triangle) than a higher power capacity with a higher uptime (yellow plus).

This trend changes significantly when considering tariffs with large demand charges (Figure 2d). In this case, there is an optimal uptime that balances the tension between reducing the overall peak power consumption (demand charge) and shifting consumption out of peak hours (energy charge). Tariffs that contain long-duration peak power charges (i.e., monthly demand charge) significantly reduce the value of flexibility. Systems with high uptime and high power capacity, however, could reap substantial benefits from demand management under select electricity tariff structures.

Facilities may also choose to participate in incentive-based demand response (IBDR) to realize additional energy flexibility benefits. While this study does not explicitly analyze participation in such programs, it is likely that high uptime and high power capacity systems will reap the largest benefits from IBDR program participation because of their high baseload power consumption coupled with low consumption during event hours.[38]

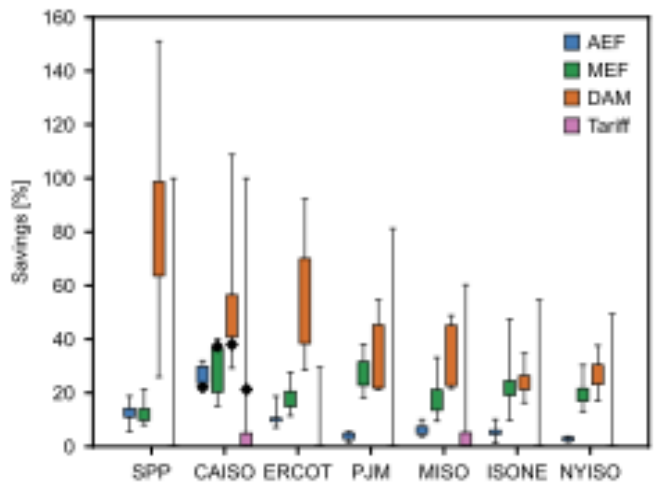

*Figure 3:* Upper bound cost and emissions benefits by region or ISO. All savings are computed relative to the base case of a 1MW time-invariant load, while the optimally flexible system schedules the same amount of energy consumed over the time horizon (24 MWh over 1 day or 744 MWh for a 31 day month) assuming minimal uptime and a PC, EC, and RTE of 100%. AEF, MEF, and DAM cases are solved for a 24-hour horizon and use signals that are averaged by month and hour of the day. The tariff cases are solved for a 1-month horizon and use all the industrial and commercial tariffs in the US large load tariff database.[37,39] Savings associated with average emissions factors (AEF) are shown in blue, marginal emissions factors (MEF) are shown in green, day-ahead market (DAM) prices are shown in yellow, and electricity tariffs are shown in pink. The variability of the box plot is derived from variation in the price signal with respect to the month of the year for AEF, MEF, and DAM, and the variation from both the month and the tariff structure for the tariff incentive. Cases that intersect with 0 have no incentive for flexibility or are completely time invariant within the 1-month timeframe. Some DAM cases have negative marginal prices during select time periods, meaning that the value of flexibility exceeds 100% of the initial cost of electricity. The black diamond markers are overlayed to represent the savings associated with July 2023 in CAISO, which is an example region and month used in Figures 2 and 4.

The upper bounds of cost and emissions savings also vary significantly by region or ISO (Figure 3). The larger magnitude in DAM and tariff savings opportunities reflect the greater variance in the underlying signal, while the greater inter-regional variance reflects the spread in underlying electricity prices and tariff structures. All incentives exhibit significant intra-region and inter-region variability. This underscores the importance of resolving spatial and temporal (seasonal and hourly) granularity when estimating benefits of flexibility at the project or case-study level. For example, systems with either a low uptime or high power capacity most effectively exploit the Spring negative pricing that occurs in SPP and CAISO. On the other hand, regions and incentives with less hourly variation, such as average emission factors (AEF) and tariffs in ERCOT and NYISO, are both more predictable and more limited in the magnitude of the energy arbitrage opportunity. Less flexible loads (high uptime and high power capacity) may be a better fit for siting in these regions and optimizing their operations around these more stable incentives.

Trade-offs between cost and emissions incentives

Past work has highlighted the temporal misalignment between electric grid emissions and costs[39] and the resulting tradeoffs when co-optimizing flexible operation in response to both emissions and price signals. We advance past work by describing the electricity cost and emissions trade-offs as a function of facility uptime and power capacity (Figure 4). Static operations (black dot) are Pareto suboptimal, except for the 25% uptime in tariffs (green) which incurs higher electricity costs than the constant-load operation (Figure 2d). This case reinforces the critical relationship between system uptime requirements and the magnitude of long duration peak power charges, such as the monthly demand charge present in the example CAISO tariff (Figure 4b and 4d).

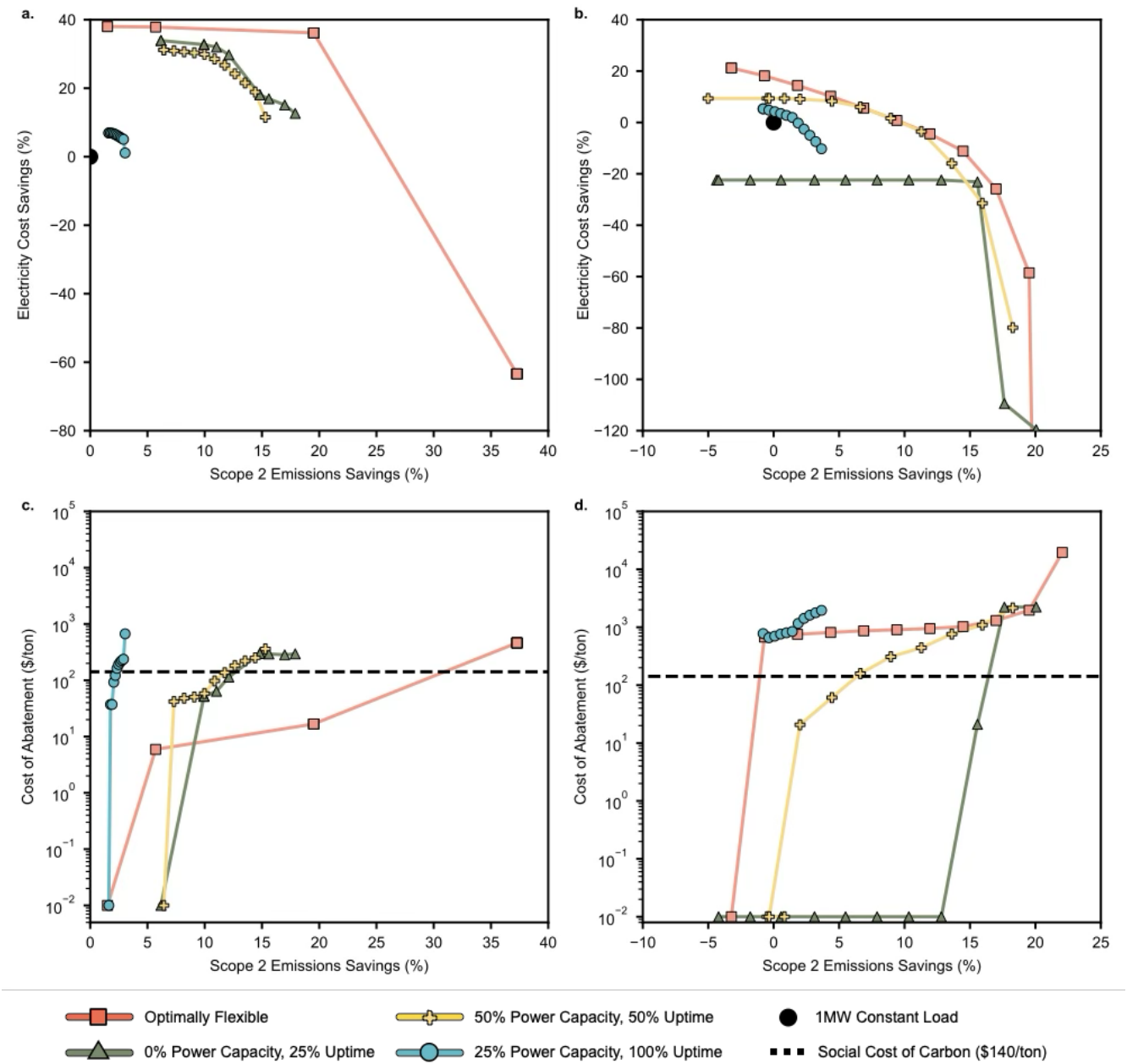

Figure 4: Trade-offs in cost and emissions savings for CAISO in July 2023 between a) marginal emissions and day-ahead market prices and b) average emissions factors and retail electricity tariff prices. The corresponding cost associated with these emissions savings is shown in c) for marginal emissions and day-ahead market prices and d) for average emissions factors and retail electricity tariff prices. Each trade-off is shown for the four facility load profiles described in Figure 1. The optimally flexible system represents the upper bound of the pareto front.

For marginal incentives (MEF and DAM), shown in Figure 4a, enhanced system flexibility increases the magnitude of the tradeoffs in flexibility for emissions, but not necessarily for costs. The optimally flexible load profile (red) has steep trade-offs at the extremes (cost optimal versus emissions optimal) because emissions optimal operation increases costs. For the three other sample systems, operations are constrained to operate over multiple timesteps, which limits achievable cost and emissions savings and their associated trade-offs. Trade-offs between AEF and tariff incentives, shown in Figure 4b, are steeper, but most emissions savings can be achieved with little to no change in cost. Achieving maximum emissions savings (red) incurs a high cost penalty due to the tariff's demand charge, but realizing 90% of the potential emissions savings increases the electricity cost by only 20%.

The cost of emissions abatement is derived as the negative of the average slope of the Pareto curve between the cost-optimal point and any other point on the curve (see Methods). For a given system and set of incentives, the cost of abatement is proportional to the percentage of emissions abated. The last 10-20% of emissions abatement via flexibility is often exponentially more costly than the first 80-90% (Figure 4). This suggests that emissions abatement from load flexibility can be tuned to balance a consumer's willingness to pay for a marginal unit of emissions abatement.

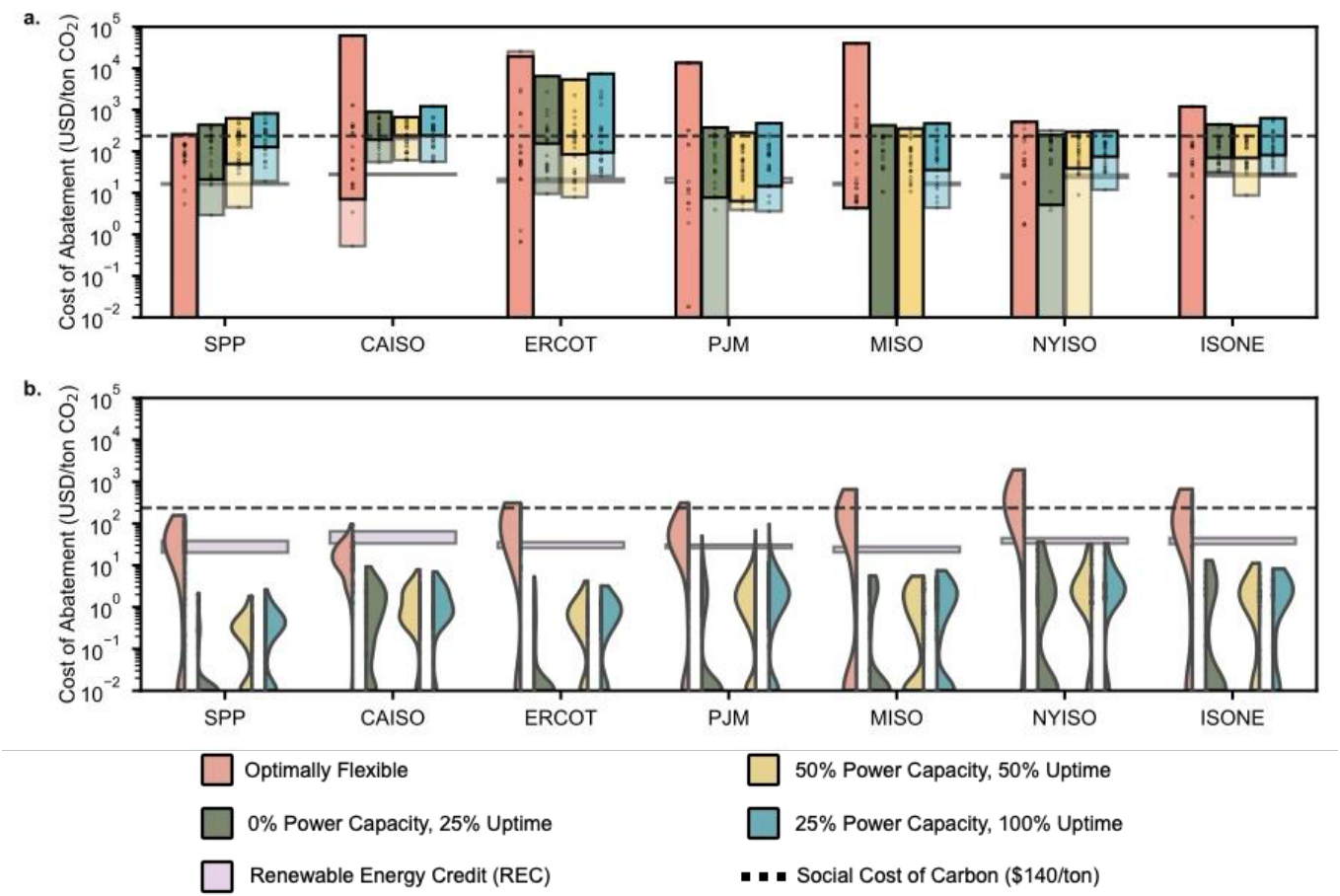

Figure 5: The cost of abatement as a function of the load's flexibility attributes when assuming 100% of potential electrical emissions abatement. The cost of abatement is calculated as the ratio of the difference in electricity cost to the electricity emissions between the cost optimal and emissions optimal operation (see Methods). The lighter bars represent the cost of abatement when the system achieves 50% of the potential electrical emissions abatement. This is visually represented as the average slope of the Pareto front in Figure 4. a) Cost of abatement for MEF and DAM signals, where the bar represents the range of cost with monthly variation and each of the points represent the cost for individual months. b) Cost of abatement for AEF and tariff signals, where the violin plots represent the combined distribution with respect to both the month and tariff. The optimally flexible case refers to the upper bound of flexibility that is also used in Figure 3. The market price for renewable energy credits are shown as a range and calculated based on the emissions intensity in each region using a nominal price of $1-20/MW (Methods).[40] The social cost of carbon ($140/ton) uses the EPA near-term rate from 2030.[41]

We compute this abatement cost as a function of load flexibility attributes and region (Figure 5). This analysis underscores the opportunity for load flexibility to deliver some degree of Scope 2 carbon abatement at prices lower than current market prices for renewable energy credits (REC) or the social cost of carbon (SCC) reference point. Complete elimination of Scope 2 emissions, however, is often significantly more costly than these benchmarks because of the relationship between the magnitude of emissions savings and the cost of abatement.

We also observe that the optimally flexible scenario often results in either extremely high or extremely low abatement costs. The optimal load profiles for single objective cost or often result in the system consuming all power in one time step (i.e., the step that results in the minimum cost or emissions). When the analysis is reformulated as a multi-objective problem, the optimal solution may advise shifting into sub-optimal load windows, depending on the degree of temporal alignment between cost and emissions profiles within the ISO region. In cases where the system has power capacity or uptime constraints, the abatement cost of emissions measures the alignment of these time-varying incentives constrained by the system's ability to exploit that arbitrage. When the cost and emissions incentives are well aligned, all forms of load flexibility can abate emissions at near zero cost. Time-invariant tariffs-based incentive structures also yield relatively low abatement costs by not financially dis-incentivizing load shifting to lower carbon windows of consumption. It should be noted that this final analysis only quantifies the value of load shifting; retail electricity costs are often multiples of wholesale electricity costs.[37]

There are many cases where load flexibility produces cheaper emissions abatement than the average REC or the SCC. RECs are modeled as abating 100% of the emissions associated with electricity in each region at a price of $1-20/MW. This is consistent with the current market for renewable energy credits.[40] However, these offsets can be uncertifiable or easily gamed, thus overvaluing this option. SCC refers to the cost of the externality of emissions and derived from the consequence of expected emissions in the long term.[41] This comparison is used as a bound on the demand for offsets, while the abatement cost shown in this study can be thought of as the cost of supplying a reduction in emissions via energy flexibility.

While the abatement cost of emissions from load flexibility may be inexpensive, easily measurable, and highly tunable, it is important to note the direct relationship between the cost and the magnitude of emissions abated. Both quantities are constrained by the dynamics of the electricity grid in the region and the system's uptime and power capacity constraints. As shown by Figure 5a, there are several cases where fully optimizing operations for emissions (dark bars) is uncompetitive with the SCC and RECs, however only abating 50% of potential emissions savings from flexibility (light bars) can cost significantly less.

**Discussion**

This study reinforces the substantial opportunity for load flexibility to reduce costs and Scope 2 carbon emissions for large industrial and commercial consumers. Incentivizing load shifting behavior requires tools for valuing flexibility amidst the heterogeneity in load shapes, spatiotemporal variation in electricity prices and emissions factors, and variability in the regional alignment of cost and emissions incentives. This study used parametric analyses of uptime, PC, EC, and RTE to inform cost and emissions-optimal design principles for load flexibility across these diverse settings. It also performed sensitivity analysis to grid region to guide investment in load flexibility and used multi-objective analysis to compare relate the value of abatement from load flexibility other market-based carbon abatement tools.

While the value of load flexibility for an individual consumer is highly specific to its load attributes (uptime, PC, EC, and RTE), location, and incentive structures, several heuristics emerged from this analysis. First, there is significant value in excess production capacity at industrial or commercial facilities (e.g., data centers, chemical manufacturing, food and beverage processing, etc.). Excess production capacity decreases the system's uptime requirement while increasing its power capacity and energy capacity, thus enabling up-flex or allowing the plant to vary short-term production while maintaining long-term targets. We also observe that the marginal costs of this excess production capacity vary significantly by industry but are often orders of magnitude lower than the cost of other energy storage mechanisms when specified during the facility design and construction phase.

Second, this work reinforces the concept that flexible loads require relatively high RTE to be cost and emissions effective. While intuitive, making this quantitative by grid region and load shape helps to inform heuristics around load flexibility for a wide variety of industries and benchmark load flexibility from virtual battery systems to that of behind the meter Li-ion batteries with RTEs upwards of 85%,[42,43] High RTE in a virtual battery system requires relatively constant efficiency over a range of operating states. This can be achieved by modularity or parallelism within a process,[3,44] using novel component design,[45] and minimizing the non-sheddable portion of the load.[46] Again, in many cases, we observe that investing in high RTE industrial processes is relatively less expensive than purchasing high RTE batteries. We also observe that for select incentive structures in select grid regions, RTEs as low as 65% may deliver value.

This study also quantifies the upper bounds of cost and emission savings by region to guide investments that maximize the value of energy flexibility. We observe significant spatial and temporal variability in the upper bounds of both cost and emissions savings. The upper bound on cost savings ranges from 0-150%, with the largest benefits available to load consumers CAISO and SPP regions that purchase from DAM markets or choose a favorable tariff structure. Consistent with previous literature, tariffs with high demand charges make operational flexibility unfavorable.[47]

In contrast, the upper bound of emissions abatement lies between 10-40%. Flexible loads in CAISO offer substantial benefits when tracking both AEF and MEF signals because average and marginal emissions factors are of the same magnitude. In other regions, however, AEFs and MEFs diverge and the MEF signal indicates much greater upper bound savings potential. For example, in ERCOT, the upper bound of emissions savings when tracking MEF can yield up to 35% savings, while tracking AEF yields only 8-20%, depending on time of year. Implementing load shifting for the explicit purpose of emissions reduction would benefit substantially from improved real-time marginal emissions estimates.[48] It should be noted that, while CAISO provides the strongest emissions incentives for flexibility, the average price of CAISO electricity is usually higher than in other regions, so the effective cost of emissions abatement may be higher.[37]

Using the bounds of savings from flexibility, we can begin to make comparisons to other cost and emissions savings measures like energy efficiency. When we compare an energy efficiency improvement of 10% for a flat load,[49] and assume the electricity cost of the system is approximately 50% of the lifetime cost, the total cost savings of production is approximately 5%. If the tariff is favorable or the consumer pays electricity that follows the DAM, investing in flexible operations may provide even greater benefit. For example, a seawater desalination plant (80% uptime and 20% power capacity) under Southern California

Edison tariffs may achieve around 18% energy cost savings from flexibility.[22] This translates to almost double the savings achievable from energy efficiency. However, in terms of gross emissions reductions, these two strategies may be similar (primarily for fully electrified loads). For individual systems there may be trade-offs between maximizing flexibility and efficiency in operation.[42]

While there is no general relationship between the maximum emissions reduction from flexibility and the cost of abatement (S4), for a given system the price of abating emissions increases exponentially with the fraction of potential emissions abated. This occurs because the potential for emissions abatement is determined by emissions intensity of generation units, while the cost of emissions abatement is set by the hourly alignment of electricity prices with emissions intensities. However, for a given system, unless the emissions and cost incentives are perfectly aligned, minimizing electricity cost results in a different trajectory of operation compared to minimizing electrical emissions.

While we find that load flexibility can provide a tunable and comparatively inexpensive mechanism for Scope 2 decarbonization, the cost of abatement has high spatiotemporal variability. The abatement cost of energy flexibility may range from $10^{-2}$ to $10^3$ \$/ton, which spans the range of unit costs associated with electrifying process heat in the food and beverage industry (\$40–70/t $CO_2$) or iron reduction in steel with carbon capture (\$90–130/t $CO_2$).[49] Importantly, these pathways are not mutually exclusive; correctly designed, electrification upgrades can increase process flexibility and yield grid co-benefits by enhancing capacity expansion potential for renewable power.

To facilitate emissions abatement via flexibility, we might consider certifying operational schedules for renewable energy credits (RECs) or incorporating emissions incentives directly into electricity tariffs. The current market for RECs and emissions offsets can be near impossible to verify. However, with power metering and grid emissions tracking, verification of emissions offsets via flexibility may soon be possible. An ongoing issue with this strategy, like the demand response markets, is certifying the baseline trajectory or counterfactual operation for power consumers.[50,51] Alternatively, energy utilities may consider incorporating emissions incentives into tariffs and selling the RECs on a separate market. If implemented correctly, this strategy could produce measurable carbon offsets in a tunable way.

**Methods**

Datasets

This study considers hourly and monthly averaged incentive data associated with marginal emissions factors (MEF), average emissions factors (AEF), day-ahead market (DAM) prices, and tariff electricity prices. The data reconciliation and time-averaging procedures and incentives are outlined in Chapin & Rao et al.[37].

Optimization formulation

To model the flexible load, we consider an optimization problem formulated to minimize the cost or emissions, over the power consumption schedule. The objective of the problem ($\phi$) is a function of power consumption ($p$) only.

$$\phi = f(p) \quad\quad\quad (1)$$

The function *f* represents the cost or emissions functions. When considering a single objective that is not a tariff, *f* is a dot product of the incentive (MEF, AEF, or DAM) and the power consumption. The objective function for tariffs is nonlinear and is built automatically using the Electric Emissions and Cost Optimizer (EECO) Python package. The power consumption is defined as non-negative with an index for each timestep. The domain may be relaxed to all real numbers if the continuous portion of power consumption is split into respective positive and negative components.

To reflect the continuous and discrete regimes of operation, the net power consumption (*p*) is a split using variables for continuous power and status.

$$p_t = p_{\text{cont},t}\, s_t \tag{2}$$

Continuous power ($p_{cont}$) is defined as a non-negative with an index for each timestep. The system status is defined as a binary variable with an index for each timestep. The sum of status variables is used to constrain the system uptime.

$$u = \frac{\sum^T s_t}{T} \tag{3}$$

The uptime (*u*) is a strictly positive variable that is defined to be less than 1. It related the timesteps where the system is on to the total number of timesteps (*T*). The average continuous power is tracked and varies based on the uptime. For a fixed uptime, the average continuous power is a linear function of the continuous power variable.

$$p_{\text{cont,avg}} = \frac{\sum^T p_{\text{cont},t}}{uT} \tag{4}$$

The power capacity ($p_c$) constrains the deviation of continuous power from the average continuous power. The power capacity is defined as a scalar parameter between 0 and 1.

$$-p_c \leq 1 - \frac{p_{\text{cont},t}}{p_{\text{cont,avg}}} \leq p_c \tag{5}$$

Conversely, the power capacity can instead be used to constrain the continuous power consumption to a predefined baseload.

$$-p_c \leq 1 - \frac{p_{\text{cont},t}}{p_{\text{base},t}} \leq p_c \tag{6}$$

The baseline power consumption ($p_{\text{base},t}$) is defined as a non-negative parameter with an index for each timestep. The domain of this parameter may be extended to real numbers if the continuous portion of power consumption is split into respective positive and negative components.

The round-trip efficiency ($\eta_{RTE}$) reflects the ratio between the baseline and the net power consumed by the flexible system. The round-trip efficiency ($\eta_{RTE}$) is a parameter defined between 0 and 1. For upper bounds calculations, the $\eta_{RTE}$ is fixed to 1.

$$\sum^T p_{\text{base},t} = \eta_{RTE} \sum^T p_{,t} \tag{7}$$

The energy capacity is calculated in post and defined as the fraction of however, it can be optionally enforced as an inequality constraint within the optimization problem. Only three out of four system characteristics (uptime, power capacity, energy capacity, and roundtrip efficiency) are required to fully define the system.

$$e_c = \frac{\sum^{\{t\,|\,p_t \leq p_{\text{base},t}\}} (p_{\text{base},t} - p_t)\Delta t}{\sum^T p_{\text{base},t}\Delta t} \qquad (8)$$

The savings associated with each incentive is calculated after optimization.

$$\% \text{ savings} = 100 * \frac{x_{\text{flexible}} - x_{\text{baseline}}}{x_{\text{baseline}}} \qquad (9)$$

Here, x refers to the incentive in consideration. For MEF, AEF, and DAM incentives, the cost savings power consumption is calculated as the dot product of the incentive array with the power consumption array produced by the model. For tariff costs, the individual charges are calculated and summed using the electric-emission-cost python package.

When the uptime is fixed, the problem is simplified to a mixed integer linear program (MILP). When the uptime is unfixed, the average continuous power constraint creates a nonlinearity. This is solved iteratively using a Lagrangian relaxation for the inequality constraint (eq. 6). After each solve, the constraint is evaluated, and the Lagrange multiplier is incremented proportional to the maximum constraint violation. The iterations are stopped when the maximum constraint violation is less than $10^{-8}$. This approach is implemented in Pyomo and solved using Gurobi 11.0.[52–54] Additional details and code associated with the solution can be found via S6.

<u>Cost and emissions trade-offs</u>

To calculate the multi-objective trade-off between cost and emissions, we solve a series of optimization problems. First, we solve the model for the cost objective alone to find the cost optimal cost and cost optimal emissions. Next, the model is solved for the emissions objective alone to find the emissions optimal cost and emissions optimal emissions. After this, the model is solved for a number of intermediate points with both cost and emissions objectives active where the emissions objective is multiplied by a Lagrange multiplier. After each solve, we calculate the electricity cost and electrical emissions.

The abatement cost of emissions is calculated as the negative of the average slope between a point on the pareto curve and the cost optimal point.

$$C_{\text{abatement}} = -\frac{c_k - c^*}{e_k - e^*} \qquad (10)$$

Here, the subscript $k$ denotes a point on the pareto curve and the superscript $*$ denotes the cost-optimal point. $C_{\text{abatement}}$ refers to the cost of abatement in units of USD/ton. c refers to the electricity cost in dollars and e refers to the electrical emissions in tons.

<u>Parameter Assumptions</u>

The social cost of carbon uses the EPA near-term rate from 2030 of $140/ton. The renewable energy credit (REC) price of $1-20/MWh is translated to $/ton by dividing $/MWh by average ton/MWh for a given region. This assumes that the REC provides an offset that fully abates the emissions associated with the electricity being used.[38] Additional details on the choice of these parameters is included in S5.

Limitations
1. Constant 1MW baseline

For simplicity and generality, this work considers a constant 1MW load as the baseline. However, this framework, and the accompanying models, can be readily applied to different baseloads including those with inherently time-varying operations.

2. The upper bound of flexibility contains no delays, zero power consumption during the offline state, and 100% RTE

We assume any process delays and base power consumption are losses in terms of flexible operation. If these exist, we assume the effects on cost can be accounted for within the round-trip efficiency term. However, more detailed models of specific processes are required to compute the effect of dynamic time delays and non-sheddable portions of the load on the net savings.

3. Power consumers are price takers and flexibility doesn't change dispatch

This assumption is common for behind-the-meter analysis. However, large loads or aggregations of loads warrant a change in the dispatch of generators from the grid operator, emissions would need to be accounted using the Adjusted Locational Marginal Carbon Emissions (ALMCE).[48]

4. Ignores ancillary services

This assumption will be effective in estimating value for most loads. A small fraction of consumers will bid into high frequency ancillary service markets (frequency regulation) or lower frequency demand response programs. This can increase the value of flexibility beyond what can be obtained from arbitrage, but it requires some characterization of the uncertainty or frequency to which these events occur.

Code availability.
All code, data, and figures associated with this paper are publicly available on GitHub: https://github.com/we3lab/flex-limits

# References

1. Graff Zivin, J. S., Kotchen, M. J. & Mansur, E. T. Spatial and temporal heterogeneity of marginal emissions: Implications for electric cars and other electricity-shifting policies. *J. Econ. Behav. Organ.* **107**, 248–268 (2014).

2. Jenkins, J. D., Luke, M. & Thernstrom, S. Getting to Zero Carbon Emissions in the Electric Power Sector. *Joule* **2**, 2498–2510 (2018).

3. Alaperä, I., Honkapuro, S. & Paananen, J. Data centers as a source of dynamic flexibility in smart girds. *Appl. Energy* **229**, 69–79 (2018).

4. Mallapragada, D. S. *et al.* Decarbonization of the chemical industry through electrification: Barriers and opportunities. *Joule* **7**, 23–41 (2023).

5. Castro, P. M., Novais, A. Q. & Carvalho, A. Optimal Equipment Allocation for High Plant Flexibility: An Industrial Case Study. *Ind. Eng. Chem. Res.* **47**, 2742–2761 (2008).

6. Liu, R. *et al.* A cross-scale framework for evaluating flexibility values of battery and fuel cell electric vehicles. *Nat. Commun.* **15**, 280 (2024).

7. Navidi, T., Gamal, A. E. & Rajagopal, R. Coordinating distributed energy resources for reliability can significantly reduce future distribution grid upgrades and peak load. *Joule* **7**, 1769–1792 (2023).

8. Park, B., Dong, J., Liu, B. & Kuruganti, T. Decarbonizing the grid: Utilizing demand-side flexibility for carbon emission reduction through locational marginal emissions in distribution networks. *Appl. Energy* **330**, 120303 (2023).

9. Xie, L. *et al.* The role of electric grid research in addressing climate change. *Nat. Clim. Change* **14**, 909–915 (2024).

10. de Sisternes, F. J., Jenkins, J. D. & Botterud, A. The value of energy storage in decarbonizing the electricity sector. *Appl. Energy* **175**, 368–379 (2016).

11. der Jagt, S. van, Patankar, N. & Jenkins, J. D. Understanding the role and design space of demand sinks in low-carbon power systems. *Energy Clim. Change* **5**, 100132 (2024).

12. Spiller, E., Esparza, R., Mohlin, K., Tapia-Ahumada, K. & Ünel, B. The role of electricity tariff design in distributed energy resource deployment. *Energy Econ.* **120**, 106500 (2023).

13. Hennig, R. J., de Vries, L. J. & Tindemans, S. H. Congestion management in electricity distribution networks: Smart tariffs, local markets and direct control. *Util. Policy* **85**, 101660 (2023).

14. Sonali Razdan, Jennifer Downing, & Louise White. *Pathways to Commercial Liftoff: Virtual Power Plants 2025 Update*. https://web.archive.org/web/20250113223720/https://liftoff.energy.gov/wp-content/uploads/2025/01/LIFTOFF_DOE_VirtualPowerPlants2025Update.pdf (2025).

15. Zhou, E. & Mai, T. *Electrification Futures Study: Operational Analysis of U.S. Power Systems with Increased Electrification and Demand-Side Flexibility*. NREL/TP--6A20-79094, 1785329, MainId:33320 https://www.osti.gov/servlets/purl/1785329/ (2021) doi:10.2172/1785329.

16. You, K. *et al.* Mitigating emissions and costs through demand-side solutions in Chinese residential buildings. *Nat. Commun.* **16**, 7358 (2025).

17. Kwac, J. & Rajagopal, R. Demand response targeting using big data analytics. in *2013 IEEE International Conference on Big Data* 683–690 (2013). doi:10.1109/BigData.2013.6691643.

18. Riepin, I., Jenkins, J. D., Swezey, D. & Brown, T. 24/7 carbon-free electricity matching accelerates adoption of advanced clean energy technologies. *Joule* **9**, (2025).

19. Jiang, W., Huber, O., Ferris, M. C. & Roald, L. Can Carbon-Aware Electric Load Shifting Reduce Emissions? An Equilibrium-Based Analysis. Preprint at https://doi.org/10.48550/arXiv.2504.07248 (2025).

20. Ricks, W., Xu, Q. & Jenkins, J. D. Minimizing emissions from grid-based hydrogen production in the United States. *Environ. Res. Lett.* **18**, 014025 (2023).

21. Puschnigg, S., Knöttner, S., Lindorfer, J. & Kienberger, T. Development of the virtual battery concept in the paper industry: Applying a dynamic life cycle assessment approach. *Sustain. Prod. Consum.* **40**, 438–457 (2023).

22. Rao, A. K., Atia, A. A., Knueven, B. & Mauter, M. S. Optimizing Desalination Operations for Energy Flexibility. *ACS Sustain. Chem. Eng.* **12**, 15696–15704 (2024).

23. Lu, M. L., Sun, Y. J., Kokogiannakis, G. & Ma, Z. J. Design of flexible energy systems for nearly/net zero energy buildings under uncertainty characteristics: A review. *Renew. Sustain. Energy Rev.* **205**, 114828 (2024).

24. Oikonomou, K. & Parvania, M. Optimal Coordination of Water Distribution Energy Flexibility With Power Systems Operation. *IEEE Trans. Smart Grid* **10**, 1101–1110 (2019).

25. Bolorinos, J., Mauter, M. S. & Rajagopal, R. Integrated Energy Flexibility Management at Wastewater Treatment Facilities. *Environ. Sci. Technol.* **57**, 18362–18371 (2023).

26. Chen, Y., Adams, T. A. I. & Barton, P. I. Optimal Design and Operation of Flexible Energy Polygeneration Systems. *Ind. Eng. Chem. Res.* **50**, 4553–4566 (2011).

27. J. Laky, D. *et al.* Market optimization and technoeconomic analysis of hydrogen-electricity coproduction systems. *Energy Environ. Sci.* **17**, 9509–9525 (2024).

28. Luke, J. *et al.* Optimal coordination of electric buses and battery storage for achieving a 24/7 carbon-free electrified fleet. *Appl. Energy* **377**, 124506 (2025).

29. Tumbalam Gooty, R. *et al.* Incorporation of market signals for the optimal design of post combustion carbon capture systems. *Appl. Energy* **337**, 120880 (2023).

30. Gjoka, K., Rismanchi, B. & Crawford, R. H. Fifth-generation district heating and cooling: Opportunities and implementation challenges in a mild climate. *Energy* **286**, 129525 (2024).

31. Wiesner, P., Behnke, I., Scheinert, D., Gontarska, K. & Thamsen, L. Let's wait awhile: how temporal workload shifting can reduce carbon emissions in the cloud. in *Proceedings of the 22nd International Middleware Conference* 260–272 (ACM, Québec city Canada, 2021). doi:10.1145/3464298.3493399.

32. Chapin, F. T., Wettermark, D., Bolorinos, J. & Mauter, M. S. Load-Shifting Strategies for Cost-Effective Emission Reductions at Wastewater Facilities. *Environ. Sci. Technol.* **59**, 2285–2294 (2025).

33. Zhou, L., Zhang, F., Wang, L. & Zhang, Q. Flexible hydrogen production source for fuel cell vehicle to reduce emission pollution and costs under the multi-objective optimization framework. *J. Clean. Prod.* **337**, 130284 (2022).

34. Li, G., Zhang, R., Bu, S., Zhang, J. & Gao, J. Probabilistic prediction-based multi-objective optimization approach for multi-energy virtual power plant. *Int. J. Electr. Power Energy Syst.* **161**, 110200 (2024).

35. Wu, W., Simpson, A. R. & Maier, H. R. Sensitivity of Optimal Tradeoffs between Cost and Greenhouse Gas Emissions for Water Distribution Systems to Electricity Tariff and Generation. *J. Water Resour. Plan. Manag.* **138**, 182–186 (2012).

36. Baumgärtner, N., Delorme, R., Hennen, M. & Bardow, A. Design of low-carbon utility systems: Exploiting time-dependent grid emissions for climate-friendly demand-side management. *Appl. Energy* **247**, 755–765 (2019).

37. Chapin, F. T. *et al.* Electricity costs and emissions incentives are misaligned for commercial and industrial power consumers. *Manuscr. Prep.* (2025).

38. Sakthivelu, A., Chapin, F. T., Bolorinos, J. & Mauter, M. S. Demand Response Event Simulator and Risk-Aware Bidding Tool for Industrial Customers. *Manuscr. Submitt. Publ.* (2025).

39. Fletcher T. Chapin, Akshay K. Rao, Adhithyan Sakthivelu, Casey S. Chen, & Meagan S. Mauter. Industrial and Commercial Electricity Tariffs in the United States. (2025).

40. Gagnon, P. *et al.* Estimating the Value of Improved Distributed Photovoltaic Adoption Forecasts for Utility Resource Planning.

41. U.S. Environmental Protection Agency. *Report on the Social Cost of Greenhouse Gases: Estimates Incorporating Recent Scientific Advances*. https://www.epa.gov/system/files/documents/2023-12/epa_scghg_2023_report_final.pdf (2023).

42. Rao, A. K., Bolorinos, J., Musabandesu, E., Chapin, F. T. & Mauter, M. S. Valuing energy flexibility from water systems. *Nat. Water* **2**, 1028–1037 (2024).

43. Hurst, K. *et al*. *Industrial Energy Storage Review*. NREL/TP--6A20-85634, 2473658, MainId:86407 https://www.osti.gov/servlets/purl/2473658/ (2024) doi:10.2172/2473658.

44. Lesniak, A., Johnsen, A. G., Rhodes, N. & Roald, L. Advanced Scheduling of Electrolyzer Modules for Grid Flexibility. Preprint at https://doi.org/10.48550/arXiv.2412.19345 (2024).

45. Johnson, H. A., Slocum, A. H. & Simon, K. P. Adaptive volutes for centrifugal pumps. (2023).

46. Rao, A., Atia, A., Bartholomew, T. & Mauter, M. Optimizing Seawater Desalination: Trade-offs in Costs and Emissions Through Flexibility. in *Optimizing Seawater Desalination: Trade-offs in Costs and Emissions Through Flexibility* vol. 23 19 (2023).

47. Rupiper, A. M., Good, R., Miller, G. J. & Loge, F. Mitigating renewables curtailment and carbon emissions in California through water sector demand flexibility. *J. Clean. Prod.* **437**, 140373 (2024).

48. Gorka, J., Rhodes, N. & Roald, L. ElectricityEmissions.jl: A Framework for the Comparison of Carbon Intensity Signals. in *Proceedings of the 16th ACM International Conference on Future and Sustainable Energy Systems* 19–30 (ACM, Rotterdam Netherlands, 2025). doi:10.1145/3679240.3734597.

49. Katheryn Scott *et al*. Pathways to Commercial Liftoff: Industrial Decarbonization. *US Department of Energy* (2023).

50. Chao, H. Demand response in wholesale electricity markets: the choice of customer baseline. *J. Regul. Econ.* **39**, 68–88 (2011).

51. Valentini, O. *et al*. Demand Response Impact Evaluation: A Review of Methods for Estimating the Customer Baseline Load. *Energies* **15**, 5259 (2022).

52. Bynum, M. L. *et al. Pyomo — Optimization Modeling in Python*. vol. 67 (Springer International Publishing, Cham, 2021).

53. Gurobi Optimization LLC. *Gurobi Optimizer Reference Manual*. https://www.gurobi.com (2023).

54. Hart, W. E., Watson, J.-P. & Woodruff, D. L. Pyomo: modeling and solving mathematical programs in Python. *Math. Program. Comput.* **3**, 219–260 (2011).

**Data Availability**

All data on emissions and tariffs used for these models are available on [GitHub](). Day-ahead market prices are obtained from Grid Status.

**Code Availability**

All models + code to generate figures is available on [GitHub]().

**Acknowledgements**

This work is supported by the National Alliance for Water Innovation (NAWI, grant number UBJQH - MSM) and the Office of Energy Efficiency and Renewable Energy (EERE, grant number 0009499 - MSM) through the Department of Energy (DOE). The authors would like to thank Kamran Tehranchi and Ines Azevedo from Stanford University for helpful conversations during the drafting process.

**Author Contributions**

**A.K.R** contributed to conceptualization, data curation, formal analysis, investigation, methodology, project administration, software, validation, and visualization.

**F.T.C.** contributed to conceptualization, data curation, formal analysis, investigation, methodology, project administration, software, validation, and visualization.

**E.M.** contributed to conceptualization, formal analysis, investigation, methodology, software, validation, and visualization.

**A.S.** Contributed to conceptualization, data curation,

**C.T.** contributed to conceptualization, data curation.

**D.W.** contributed to conceptualization, data curation, visualization.

**M.S.M.** contributed to conceptualization, funding acquisition, project administration, supervision, and validation.

All authors contributed to the drafting and writing process.

**Supplemental Information:** "How much can we save? Upper bound cost and emissions benefits from commercial and industrial load flexibility"

Akshay K. Rao[a], Fletcher T. Chapin[a], Erin Musabandesu[a], Adhithyan Sakthivelu[a], Carson Tucker[a], Daly Wettermark[a], Meagan S. Mauter[a,b,c,d,e †]

[a] Department of Civil and Environmental Engineering, Stanford University, 473 Via Ortega, Stanford, California 94305, United States
[b] Environmental Social Sciences, Stanford University, 473 Via Ortega, Stanford, California 94305, United States
[c] Senior Fellow, Woods Institute for the Environment, Stanford University, 473 Via Ortega, Stanford, California 94305, United States
[d] Senior Fellow, Precourt Institute for Energy, Stanford University, 473 Via Ortega, Stanford, California 94305, United States
[e] Photon Science, SLAC National Accelerator Laboratory, 2575 Sand Hill Road, Menlo Park, California 94025, United States

*†Corresponding author*: mauter@stanford.edu

**S1.** Upper bound emissions and cost savings

**S2.** Energy capacity as a function of system flexibility

**S3**. Sensitivity of savings to RTE

**S4**. Cost of Abatement trends across ISOs

**S5**. Calculating Renewable Energy Credit Value

**S6.** Formulation of the optimization problem

**S7.** Additional references

**S1.** Upper bound emissions and cost savings

The main text Figure 2 features the upper bound emissions and cost savings as a function of a CAISO load consumer's uptime and power capacity in July.

Similar figures for all ISOs and months can be found on github: https://github.com/we3lab/flex-limits/tree/main/paper_figures/figures/pdf/max_savings_contours

**S2.** Energy capacity as a function of system flexibility

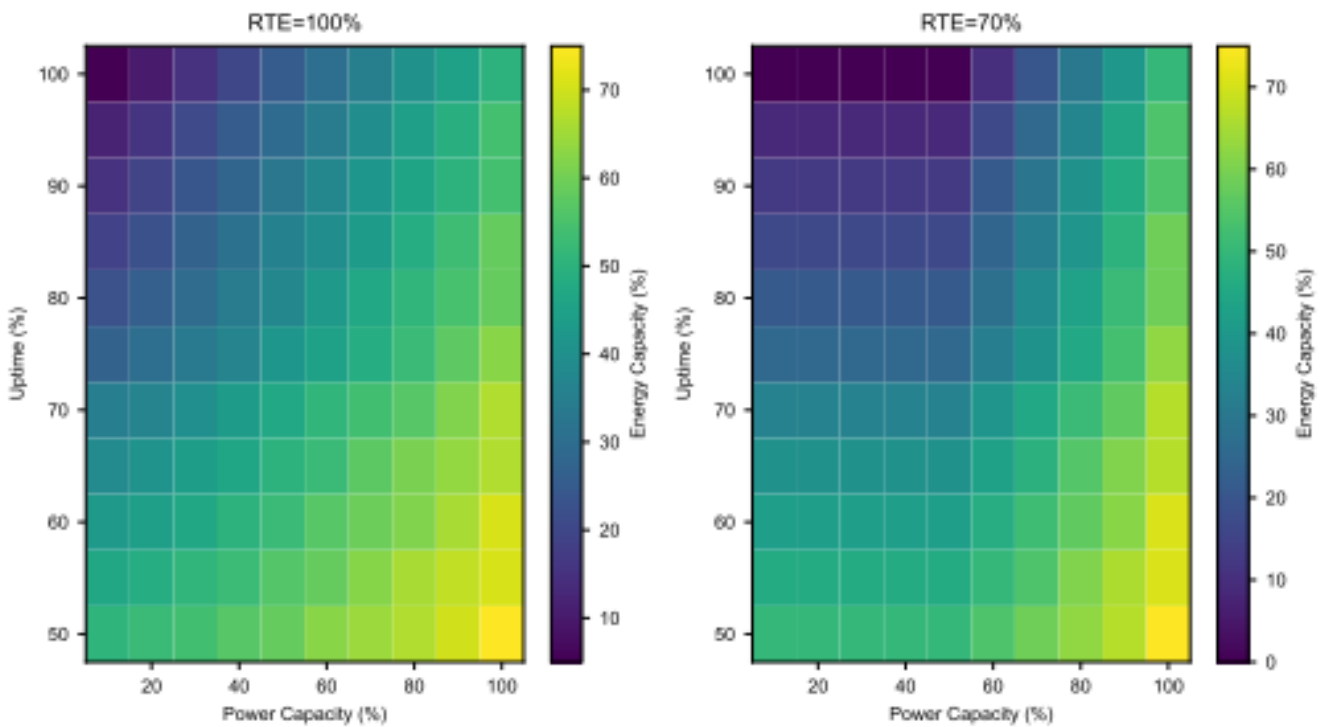

Figure S1. The energy capacity is shown across the space of uptime and power capacity for two round-trip efficiency values (100% left, 70% right). This is calculated for day-ahead market prices in SPP in January, but there is small variation across incentives (MEF, AEF) and months.

Code to reconstruct this contour map for other regions and months can be found on GitHub: https://github.com/we3lab/flex-limits/blob/main/paper_figures/code/energy_capacity_analysis.py

**S3.** Sensitivity of savings to RTE

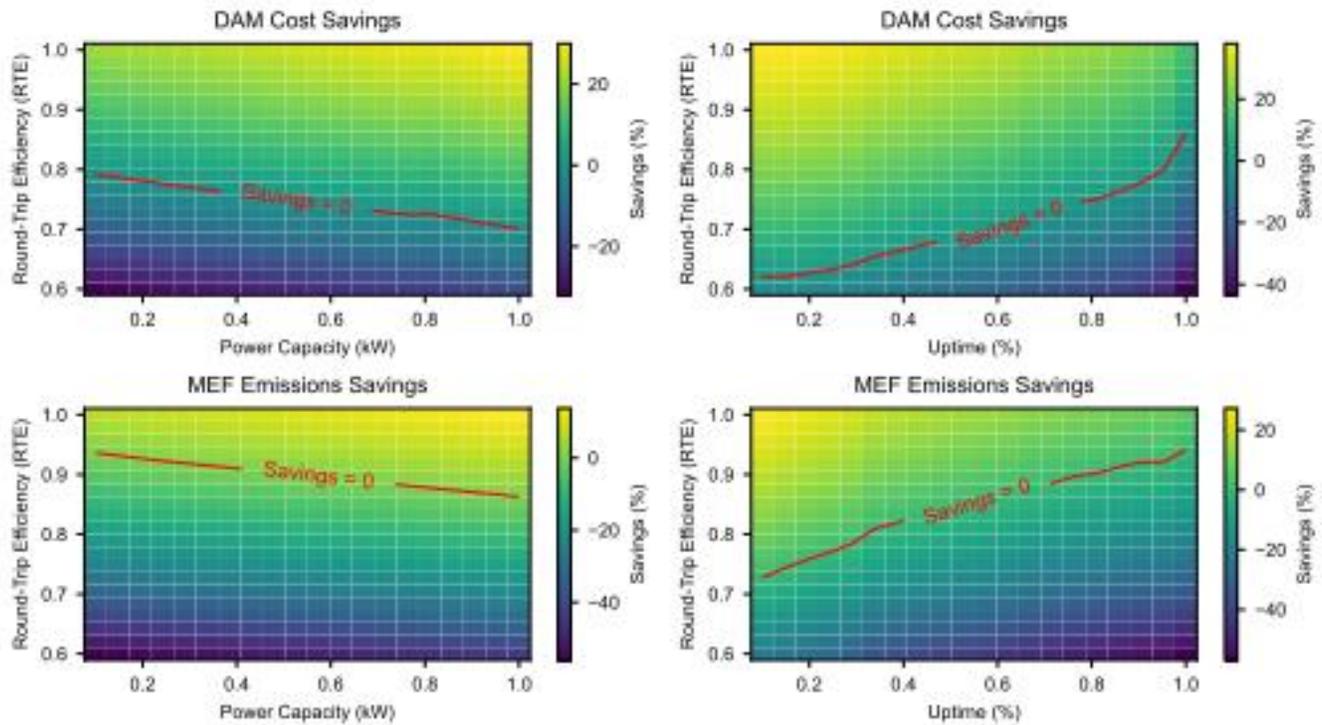

Figure S2. For RTE< 1, savings in day ahead market pricing (a, b) and marginal emissions (c, d) in CAISO in July. For this example's purpose, cases with varying power capacity assume an 80% uptime and for cases with varying uptime we assume a 50% power capacity.

While the preceding analysis was performed for systems with 100% EC and RTE, EC and RTE can vary substantially. We find that most systems need at least 60%-99% RTE to show any benefit from flexibility. This minimum RTE is largely dependent on the magnitude of the upper bound of savings.

Code to reconstruct these contour maps for other regions and months can be found on GitHub: https://github.com/we3lab/flex-limits/blob/main/paper_figures/code/rte_analysis.py

**S4**. Cost of Abatement trends

The cost of abatement is calculated by comparing the difference in cost divided by the difference in emissions between two points on the pareto trade-off curve. Figures 4 and 5 of the main text describe the cost of abatement for four example systems, however this is defined for the full space of uptime and power capacity. There is no discernable trend across the space of uptime and power capacity that persists across ISOs and months.

Figures and data for all cases can be found on GitHub: https://github.com/we3lab/flex-limits/tree/main/paper_figures/figures/pdf/shadowcost_contour_mef_dam_abatement1.0

**S5**. Calculating Renewable Energy Credit Value

We use the following approach to compare the abatement cost of carbon for energy flexibility and with the Social Cost of Carbon (SCC) and representative values for Renewable Energy Credits (RECs).

The SCC aims to quantify the externality cost of today's greenhouse gas emissions on the future economy and social well-being. Because it considers future costs, EPA's published values for SCC include a range discount rates. We select the middle of the three near-term discount rates, 2.0%, which aligns with the White House Office of Management and Budget's guidance on regulatory analysis.[1]

We determine the average hourly REC-equivalent abatement cost for each hour of the day during each month. We divide the REC price by the hourly average emissions factor for each unique combination of month and hour of the day in 2023. This is shown below in (1) for marginal emissions factors (MEFs) but also applied to average emissions factors (AEFs) where appropriate.

$$REC\ Abatement\ Cost_{hour}\left[\frac{\$}{ton\ CO_2}\right] = \frac{REC\ Prices\left[\frac{\$}{MWh}\right]}{\frac{\sum_{days}^{D} MEF_{hour,day}\left[\frac{ton\ CO_2}{MWh}\right]}{\#\ days} \times} \quad (1)$$

REC values are priced per Megawatt-Hour of electricity delivered ($/MWh). For analysis, we assume a range of REC prices from $1-20/MW.[2] To compare to the abatement cost of carbon for energy flexibility, which is expressed in $/kg CO2-e, requires determining the approximate mass of avoided $CO_2$-equivalent emissions. To convert the REC price to an estimated CO2-equivalent, we assume that the REC is generated from zero-carbon electricity. While renewable energy does have a small amount of embodied emissions, this assumption is conservative in making claims about the value of flexibility, as it is favorable to RECs in comparison

The actual cost of RECs can vary significantly in cost and quality based on geographic location due to local renewable portfolio standards (RPS) and demand for renewable energy. Beyond the compliance-based RECs that are used in this study, there are also voluntary RECs and Solar RECs (SRECs). Voluntary RECs are more likely purchased by public or private entities who have set their own sustainability goals and may have varied degrees of quality (verifiability and emissions abatement). Solar RECs (SRECs) are a sub-type of high-value compliance REC, driven by RPS requiring a minimum fraction of solar. Electricity consumers on retail rates (i.e., tariffs) may only purchase voluntary RECs, as RPS requirements are not relevant to retail customers.

While most REC price data is not publicly available, as it is owned by private entities, we source examples of REC prices from a 2018 NREL study giving median REC prices. We compiled additional REC example prices from industry articles published within 2 years of the study period (2023). This data is included in our codebase but not represented in figures. These include nationwide typical values for 2022 were included in a newsletter from S&P Global Market Intelligence and some ISO-specific values from sub-regions.

- CAISO: Two examples published online by utilities[3],[4]
- PJM: 7 state-specific SREC examples published for compliance SREC values[5]

For remaining regions where typical REC prices could not be found, we use the nationwide typical values for 2022 were included in a newsletter from S&P Global Market Intelligence[6]

**S6**. Formulation of the optimization problem

The mixed integer optimization program minimizes the multi-objective cost and emissions subject to load constraints that tie the flexible load operation to the baseline by the uptime (number of timesteps it is allowed to be online), the power capacity (the maximum deviation from the baseload power consumption), and the round-trip efficiency (the ratio of total energy consumed between the flexibly operating system relative to the baseload).

When the uptime is treated as an equality constraint, the number of timesteps where the system is online is fixed and therefore the upper and lower bounds of continuous power consumption are also fixed. This form of the problem becomes a mixed integer linear program. This model is solved with Gurobi.

When the uptime is treated as an inequality constraint (i.e., minimum uptime), the number of timesteps where the system is unknown. This becomes a mixed integer nonlinear program because the bounds of power capacity change with the amount of time where the system is online. To solve this, we apply a Lagrange relaxation to the power capacity bounds by removing the hard constraint and instead penalizing the objective function with violations of this constraint. This is solved iteratively by incrementing the duals proportional to the violation of this soft constraint until convergence with a tolerance of 1e-8.

This model can be found on GitHub: https://github.com/we3lab/flex-limits/blob/main/models/flexload_milp.py

**S7.** Additional References

1. Office of Management and Budget. Circular No. A-4. (2023).

2. Gagnon, P. *et al*. Estimating the Value of Improved Distributed Photovoltaic Adoption Forecasts for Utility Resource Planning.

3. Silicon Valley Power. Large Customer Renewable Energy FAQs. *Silicon Valley Power* https://www.siliconvalleypower.com/sustainability/large-customer-renewable-energy-lcre-program/large-customer-renewable-energy-faqs (2023).

4. Southern California Edison. Renewable Energy Credit (REC). (2023).

5. Solar.com. SRECs: What are Solar Renewable Energy Credits? *Solar Learning Center* https://www.solar.com/learn/what-are-solar-srecs/ (2023).

6. Wilson, A. & Lenoir, T. US renewable energy credit market size to double to $26 billion by 2030. *Market Intelligence* https://www.spglobal.com/market-intelligence/en/news-insights/research/us-renewable-energy-credit-market-size-to-double-to-26-billion-by-2030 (2022).